\documentclass[aps, prb, twocolumn, superscriptaddress, amsmath,  tightenlines, longbibliography]{revtex4-1}

\usepackage{dcolumn}
\usepackage{graphicx}
\usepackage{mathrsfs}
\usepackage{subfigure}
\usepackage{booktabs}
\usepackage{amsmath}
\usepackage{physics}
\usepackage{dsfont}
\usepackage{amstext}
\usepackage{amssymb}
\usepackage{amsbsy}
\usepackage{bbm}
\usepackage{amsthm}
\usepackage{graphicx}
\usepackage{color}
\usepackage[colorlinks,citecolor=blue]{hyperref}

\usepackage[left=1.3cm,   right=1.3cm,
top=1.20cm,    bottom=1.20cm]{geometry}

\usepackage{url}
\usepackage[colorlinks]{hyperref}
\hypersetup{%
	plainpages=true,
	breaklinks=true,       
	hypertexnames=false,  
	pageanchor=true,
	colorlinks=true,
	linkcolor={blue},
	citecolor={red},
	urlcolor={blue},
	anchorcolor={black}
}

 \makeatletter

\newcommand{\Rmnum}[1]{\expandafter\@slowromancap\romannumeral #1@}
\makeatother

\begin{document}

\title{Observation of Impurity-Induced Scale-Free Localization  \\ in a Disordered Non-Hermitian Electrical Circuit }
\author{Hao Wang}
\thanks{These authors contributed equally}
\affiliation{School of Physics and Optoelectronics, South China University of Technology,  Guangzhou 510640, China}
\author{Jin Liu}
\thanks{These authors contributed equally}
\affiliation{School of Physics and Optoelectronics, South China University of Technology,  Guangzhou 510640, China}
\author{Tao Liu}
\email[E-mail: ]{liutao0716@scut.edu.cn}
\affiliation{School of Physics and Optoelectronics, South China University of Technology,  Guangzhou 510640, China}
\author{Wen-Bo Ju}
\email[E-mail: ]{juwenbo@scut.edu.cn}
\affiliation{School of Physics and Optoelectronics, South China University of Technology,  Guangzhou 510640, China}

\date{{\small \today}}


\begin{abstract}	
One of unique features of  non-Hermitian systems  is the extreme sensitive to their boundary conditions, e.g., the emergence of non-Hermitian skin effect (NHSE) under the open boundary conditions, where most of bulk states become localized at the boundaries. In the presence of impurities, the scale-free localization can appear, which is qualitatively distinct from the NHSE. Here, we experimentally design a disordered non-Hermitian electrical circuits in the presence of a single non-Hermitian impurity and the nonreciprocal hopping. We observe the anomalous scale-free accumulation of  eigenstates, opposite to the bulk hopping direction. The experimental results open the door to further explore the anomalous skin effects in non-Hermitian electrical circuits.
\end{abstract}

\maketitle

\section{Introduction} 

Growing efforts have been invested to intriguing phenomenon of non-Hermitian systems in recent year \cite{Ashida2020,PhysRevLett.116.133903, PhysRevLett.118.040401, PhysRevLett.118.045701,arXiv:1802.07964,El-Ganainy2018,ShunyuYao2018,PhysRevLett.125.126402,PhysRevLett.123.066404, YaoarXiv:1804.04672,PhysRevLett.121.026808,PhysRevLett.122.076801,PhysRevLett.123.170401,Zhang2021, PhysRevLett.123.206404,PhysRevLett.123.066405,PhysRevLett.123.206404,Sun2021, PhysRevB.100.054105,PhysRevB.99.235112,Zhao2019,PhysRevX.9.041015,PhysRevLett.124.056802,PhysRevB.102.235151, Li2020,PhysRevB.104.165117,    PhysRevLett.124.086801,Zou2023,Fan2021, PhysRevLett.125.186802,PhysRevLett.127.196801,Li2021, RevModPhys.93.015005,PhysRevLett.128.223903,  Zhang2022,Lin2023, Ren2022,PhysRevX.13.021007,PhysRevLett.131.036402,PhysRevLett.131.116601,arXiv:2311.03777,arXiv:2311.06550,arXiv:2403.07459,arXiv:2311.06550,arXiv:2311.03777,arXiv:2403.07459,PhysRevLett.132.050402,PhysRevX.14.021011,Xie2024,Zhang2024}. One of unique features in non-Hermitian systems is the non-Hermitian skin effect (NHSE) \cite{ ShunyuYao2018,PhysRevLett.125.126402,PhysRevLett.123.066404, YaoarXiv:1804.04672,PhysRevLett.121.026808,PhysRevLett.122.076801,PhysRevLett.123.170401}. This effect is characterized by an extreme sensitivity of  eigenspectra to boundary conditions, where most of bulk modes become localized at the boundaries under open boundary conditions (OBCs).  A lot of exciting non-Hermitian phenomena without their Hermitian counterparts are related to the NHSE, e.g., breakdown of conventional Bloch band theory \cite{ShunyuYao2018},  scale-free localization \cite{Li2021}, and disorder-free entanglement phase transitions \cite{PhysRevX.13.021007}. The NHSEs have been experimentally observed in many physical systems and have also shown potential applications in sensors due to the extreme sensitivity to the boundary conditions \cite{PhysRevResearch.4.013113,PhysRevLett.125.180403,McDonald2020} .

For NHSE, the localization length of bulk modes is usually independent of the system's size under OBCs. Recently, an anomalous skin localization, dubbed  scale-free localization, was found and extensively explored in non-Hermitian system \cite{Li2021,PhysRevB.107.134121,PhysRevB.108.L161409,Wang2023,PhysRevB.104.165117,PhysRevB.108.205423,PhysRevB.109.L140102,PhysRevResearch.5.033058}.   Unlike the conventional NHSE, the localization length of scale-free modes relies on the system size, and the localization direction is not indicated by the bulk. This intriguing localization phenomenon has been largely investigated in various non-Hermitian systems \cite{Li2021,PhysRevB.107.134121,PhysRevB.108.L161409,Wang2023,PhysRevB.104.165117,PhysRevB.108.205423,PhysRevB.109.L140102,PhysRevResearch.5.033058}. Recently, the scale-free localization has been experimentally observed in an electrical circuit with a Hermitian lattice subjected to a parity-time-symmetric non-Hermitian defect \cite{PhysRevB.109.L140102}. While, the experimental observation of the scale-free localization, resulting from the interplay of nonreciprocal hopping in the bulk and the single impurity, is still lacking.

An electrical circuit has become a powerful platform to
realize topological structures even with complicated lattice geometries, e.g., higher-order topological Anderson insulator,  novel topological states in hyperbolic lattices, and among others \cite{PhysRevLett.126.146802,PhysRevLett.130.206401,Imhof2018,PhysRevLett.123.053902,PhysRevB.99.020304,Zhang2023,Zhang2022ob}. Due to the design flexibility, the nonreciprocal hopping can be easily realized by using  operational amplifiers arranged as impedance converters through current inversion (INIC) \cite{PhysRevLett.122.247702}. Therefore, the electrical circuits have been utilized to realize    novel non-Hermitian phenomena  \cite{Lee2018,PhysRevB.99.161114,PhysRevLett.122.247702,PhysRevB.100.184202,Wang2020,PhysRevB.100.201406,PhysRevB.99.020302}.     In this work, we experimentally designed the non-Hermitian electrical circuit in the presence of the nonreciprocal hopping, a single non-Hermitian impurity and onsite disorder. We measure and observe the scale-free localization in the disordered non-Hermitian chain. Such  anomalous scale-free accumulations of  eigenstates are controlled by the single non-Hermitian impurity, and their localization direction can be opposite to the bulk hopping direction. Our experiment verifies the existence of the anomalous skin effects induced by the single impurity in the nonreciprocal systems, and  the results open the door to further explore the interesting localization phenomena in non-Hermitian electrical circuits.

\begin{figure*}[!tb] 
	\centering
	\includegraphics[width=19cm]{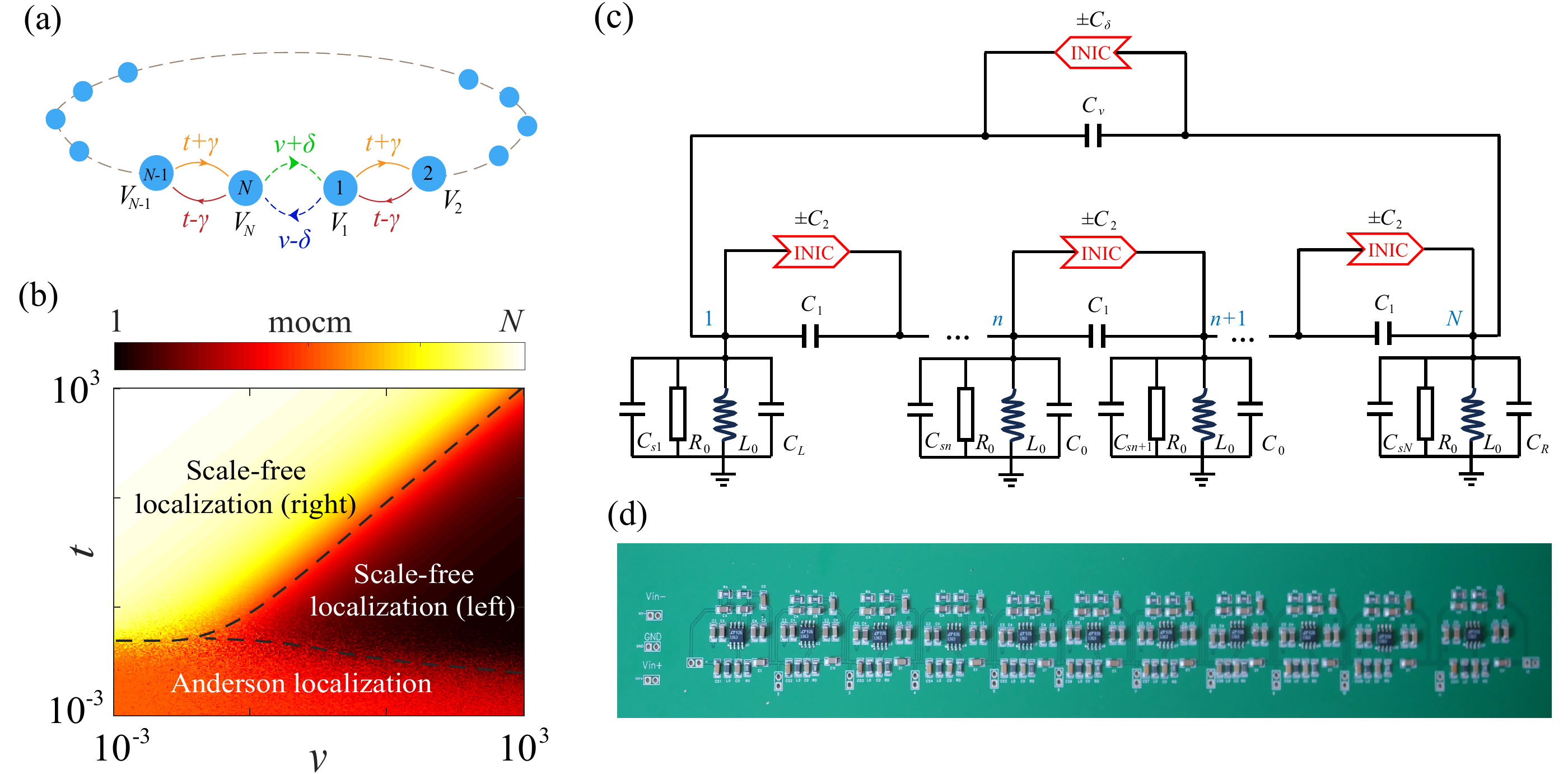}
	\caption{ (a) Schematic of HN model in the presence of onsite disorder and a single non-Hermitian impurity. $t\pm \gamma$ denotes  the nonreciprocal hopping strength, $v\pm \delta$ is the nonreciprocal hopping strength between the first and last sites, severing as a single non-Hermitian impurity, and $V_n\in[-V,~V]$ is the random onsite potential with $V=0.05$.  (b) Phase diagram of the model as  functions of $t$ and $v$ with $t= \gamma$ and $v=\delta \neq 0$.  (c) Electrical circuit implementation of the model. The nodes are interconnected by INIC and capacitors in parallel, achieving nonreciprocal hopping. (d) Photographne of the experimental circuit board.}\label{fig1}
\end{figure*}

\section{Model and Non-Hermitian electrical circuit}

In order to study the anomalous skin effect due to the interplay of disorder and impurity, we consider the disordered Hatano-Nelson (HN) chain in the presence of a single non-Hermitian impurity, with its Hamiltonian reading  \cite{PhysRevResearch.5.033058}
\begin{align}\label{H1}
	\mathcal{H}  = & \sum _ {n=1}^ {N-1} \left[(t+ \gamma )\ket{n+1}\bra{n}+(t- \gamma )\ket{n}\bra{n+1} \right] \nonumber \\ & + \sum _ {n=1}^ {N} V_ {n} \ket{n}\bra{n}+(v+\delta)\ket{1}\bra{N}+(v-\delta)\ket{N}\bra{1},
\end{align}
where $t\pm \gamma$ indicate the asymmetric hopping strengths, $V_n$ is onsite disorder potential, sampled in a random uniform distribution $[-V,~V]$, and $v \pm \delta$ are the asymmetric hopping strengths between the first and last sites, severing as a single non-Hermitian impurity [see Fig.~\ref{fig1}(a)]. By controlling the impurity's parameters $v \pm\delta$ in the presence of the disorder, one can observe anomalous skin-localization phenomena  \cite{PhysRevResearch.5.033058}, where the system undergoes  Anderson localization and scale-free skin localization, as indicated by the phase diagram in Fig.~\ref{fig1}(b). The phase diagram is obtained by calculating the mean center of mass (mcom), which is defined as the amplitude squared of all right eigenvectors $\ket{\psi_{R,n}}$, averaged over many disorder realizations $N_r$ \cite{PhysRevResearch.5.033058}, i.e.,
\begin{align}\label{eq:mcom}
	\mathrm{mcom}  = \frac{\sum_{j=1}^N j \left< \mathcal{A}(j) \right>_V}{\sum_{j=1}^N \left< \mathcal{A}(j) \right>_V},  
\end{align}
with
\begin{align}
	\left< \mathcal{A}(j) \right>_V &= \left< \frac{1}{N} \sum_{n=1}^N  (|\bra{{\color{red}j}} \ket{\psi_{R,n}}|^2  \right>_V.
\end{align}
Here, $\left< \cdot \right>_V$ indicates disorder averages.

The nonreciprocal hopping typically leads to non-Hermitian skin effects in the clean system. While, the scale-free localization in the presence of a single non-Hermitian impurity is distinct from the non-Hermitian skin effect, where its localization is not dictated by the bulk, and the localization length is proportional to the system size \cite{PhysRevResearch.5.033058}. Furthermore, the localization at the left or right boundary of the chain is controlled by the impurity hopping strength in the  1D disordered HN chain  [see Fig.~\ref{fig1}(b)].  Note that such anomalous skin-localization feature is determined by the interplay of bulk hopping strength and the single non-Hermitian impurity, but it is stabilized to a nonmonotonic localization behavior as a function of the hopping terms by the random disorder \cite{PhysRevResearch.5.033058}, as shown in Fig.~\ref{fig1}(b).

\begin{figure*}[!tb]
	\centering
	\includegraphics[width=19cm]{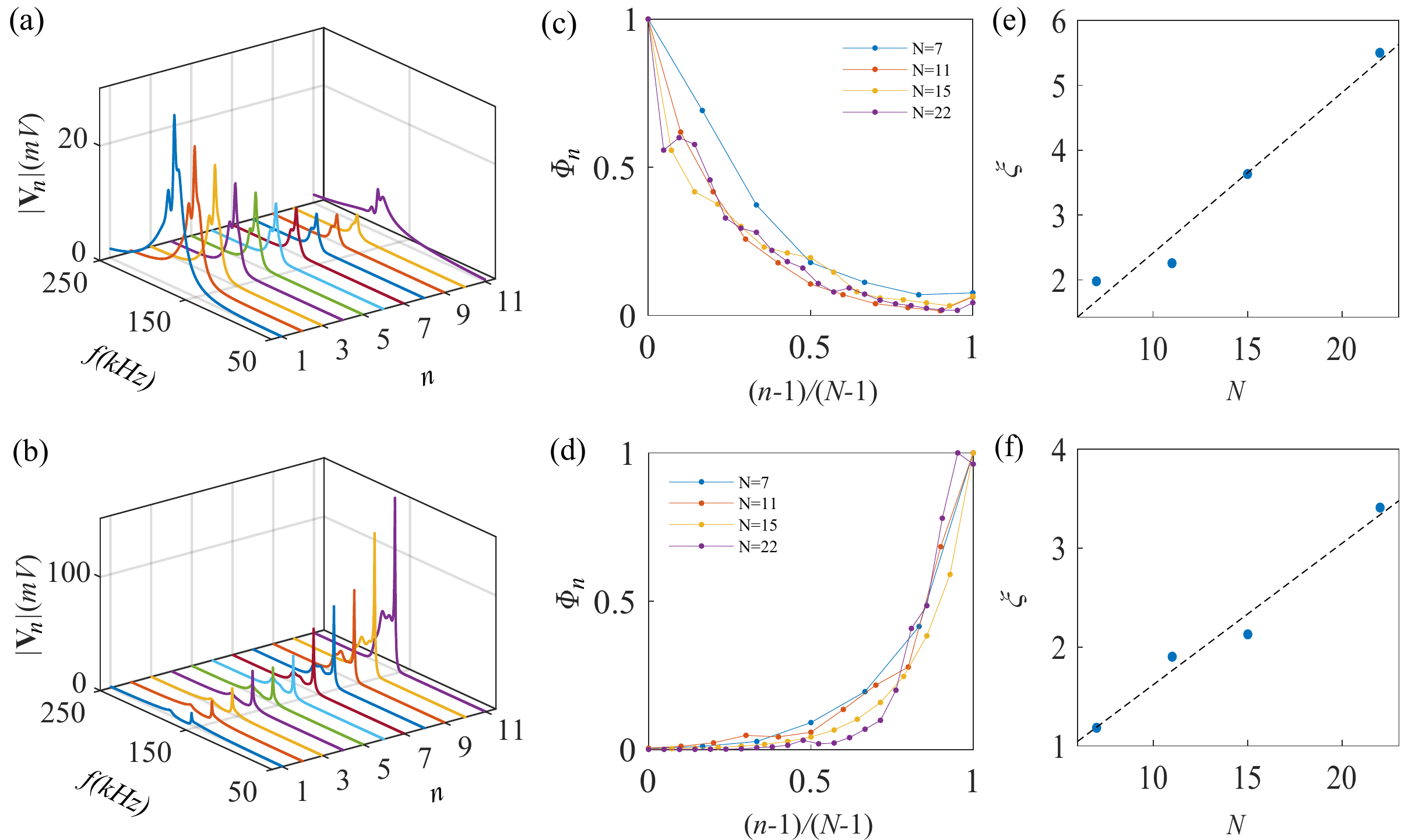}
	\caption{ Simulated results for the scale-free localization in the electrical circuit. Frequency-resolved voltage distribution $\abs{\mathbf{V}_n(\omega)}$  excited by the alternating current (AC) at the different node  $n$ (a) for  $C_1 = 9.4$ nF and $C_v = 47$ nF, and (b) for  $C_1 = 22$nF and $C_v = 2.2$ nF. (c, d) The corresponding normalized spatial distribution $\Phi_n$ of  the voltage at different normalized node indices $(n-1)/(N-1)$ for different lattice size $N$, where the node index  is mapped to the range $[0,~1]$.  (e, f) Localization length $\xi$ (blue dots) of bulk modes as a function of the lattice size $N$. The black dashed line denotes a linear fit to $\xi$. }\label{fig2}
\end{figure*}

\begin{figure}[!tb]
	\centering
	\includegraphics[width=9.2cm]{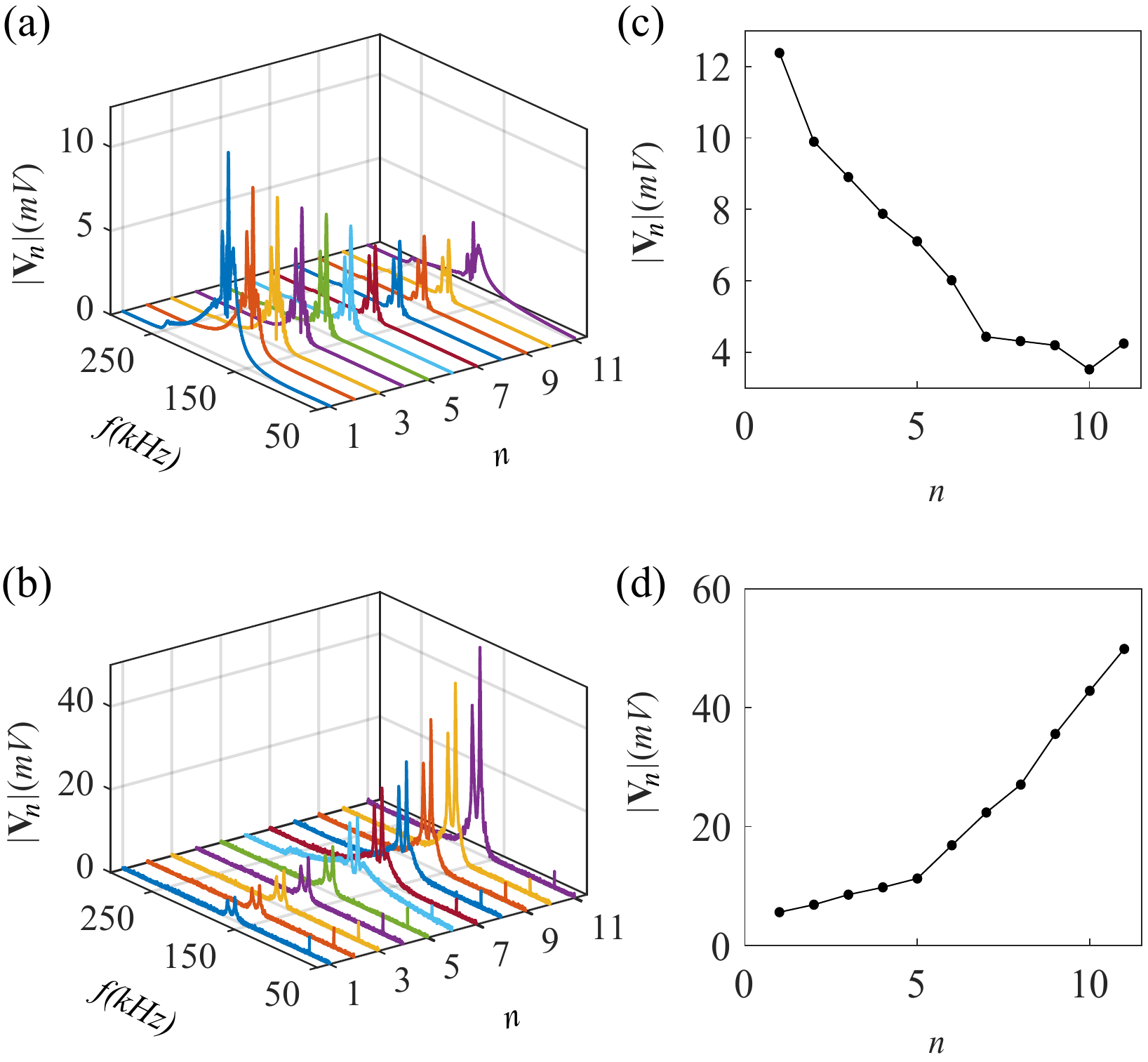}
	\caption{ Experimentally measured voltages of the admittance under chirp signal excitation. Frequency-resolved voltage distribution (a) for  $C_1 = 9.4$nF and $C_v = 47$nF, and (b) for  $C_1 = 22$nF and $C_v = 2.2$nF, respectively. (c,d) The corresponding spatial distribution of the voltage at the peak frequency, indicating the left-localized and right-localized states.}\label{fig3}
\end{figure}

In order to experimentally observe the anomalous skin-localization phenomena due to the interplay of a single non-Hermitian impurity and the bulk nonreciprocal hopping, we design non-Hermitian electrical circuits, corresponding to the model in Eq.~(\ref{H1}). Figure \ref{fig1}(c) plots the electrical circuit network, where the nonreciprocal hopping between nodes $n$ and $n+1$ is realized by the negative impedance converters through current inversions (INICs) \cite{PhysRevLett.122.247702}. Figure \ref{fig1}(d) shows the experimental circuit board, where the first node and the last node are connected by the external wires acting as the single non-Hermitian impurity.  The disorder term $V_ {n}$ in Eq.~(\ref{H1}) is introduced by the grounded capacitor $C_{sn}$ ($n=1,2,\cdots, N$) and the tolerance of the grounded inductance $L_0$ [see Fig.~\ref{fig1}(c)].  The model in Eq.~(\ref{H1}) is represented by the circuit Laplacian $J(\omega)$ of the circuit \cite{Lee2018}. The Laplacian is defined as the grounded-voltage vector  $\mathbf{V}$ to the vector $\mathbf{I}$ of input current   by $\mathbf{I}(\omega) = J(\omega)\mathbf{V}(\omega)$. As shown in Fig.~\ref{fig1}(c), the circuit Laplacian reads (see Appendix \ref{AppendixA})
\begin{align}\label{admittence}
	J = ~ & i \omega \sum _ {{\color{red}n=1}}^ {N-1} (- C_{2}- C_{1})\ket{n+1}\bra{n}+(C_{2} - C_{1} )\ket{n}\bra{n+1}  \nonumber \\ 
	& +(- C_{\delta} -C_{v} )\ket{1}\bra{N}+( C_{\delta} -C_{v})\ket{N}\bra{1}\nonumber \\ & +i \omega\sum _ {{\color{red}n=1}}^ {n} \left[C_{sn}-\frac{C_S}{2} -\varepsilon(\omega)\right]\ket{n}\bra{n},
\end{align}
with
\begin{align}\label{admittence2}
	\varepsilon(\omega)=\frac{1}{ \omega^2 L_0} -2C_1-C_0-\frac{C_S}{2}+\frac{i}{\omega R_0},
\end{align}
where   $C_{sn}$ signifies a grounded capacitor at the node $n$ {\color{red}within} the range $[0,~C_S]$.  By further writing $J$ as $J=i\omega[\mathcal{H}-\varepsilon(\omega)]$, one found that  $J$ and $\mathcal{H}$ share the same eigenstates, if we set $\pm C_2-C_1 = t \pm \gamma$, $\pm C_\delta-C_v = v \pm \delta$, and $C_{sn}- C_S/2  = V_n$. The eigenvalues and eigenstates of $J$ can be obtained by measuring the voltage response at the circuit nodes.

\section{Electrical-circuit simulation of scale-free  localization}

It has shown that the single non-Hermitian impurity can induce a scale-free accumulation of all eigenstates opposite to the bulk hopping direction [see Fig.~\ref{fig1}(b)], distinct from the NHSE occurring at open boundaries \cite{PhysRevResearch.5.033058}. Such scale-free localization phenomenon is simulated using electrical circuit, as shown in Fig.~\ref{fig2}. Here, we set $C_1 = C_2$, $C_v = C_\delta$, and introduced onsite disorder through random variations in the fabricated grounded inductors due to imperfect manufacturing processes. 

The voltage distribution at resonance frequency can be used to represent the state distribution of the circuit Laplacian. Figure \ref{fig2}(a,b)  plots the frequency-resolved voltage distribution $\abs{\mathbf{V}_n(\omega)}$ excited by alternating current (AC) at the different node $n$ for  $C_1 = 9.4$ nF and $C_v = 47$ nF, and (b) for  $C_1 = 22$ nF and $C_v = 2.2$ nF,  corresponding to the skin-mode localized at the left and right sides of the chain, respectively. This indicates that the bulk states localized towards different directions can be controlled by changing the hopping strength  within the bulk chain and the the hopping strength at the single-impurity site in spite of the nonreciprocal hopping direction within the bulk.

Figure  \ref{fig1}(b) shows the existence of the scale-free localization controlled by the single non-Hermitian impurity in the presence of weak disorder, where the localization length is  dependent on the lattice size. In order to demonstrate the scale-free localization, we calculate the  normalized  spatial distribution $\Phi_n$ of  the peak voltage at the node $n$, which is defined as
\begin{align}\label{H2}
	\Phi_n  =  \frac{ \mathbf{V}_n^2 |_{\textrm{peak}} }{\text{max}[\{ \mathbf{V}_1^2 |_{\textrm{peak}}, ~\mathbf{V}_2^2|_{\textrm{peak}},~\cdots, ~\mathbf{V}_N^2|_{\textrm{peak}}\}]}, 
\end{align}
where $\mathbf{V}_n^2 |_{\textrm{peak}} $ denotes the peak voltage at node $n$, which is normalized to the maximum value of   peak voltages of all the nodes.

\begin{figure*}[!tb]
	\centering
	\includegraphics[width=17.4cm]{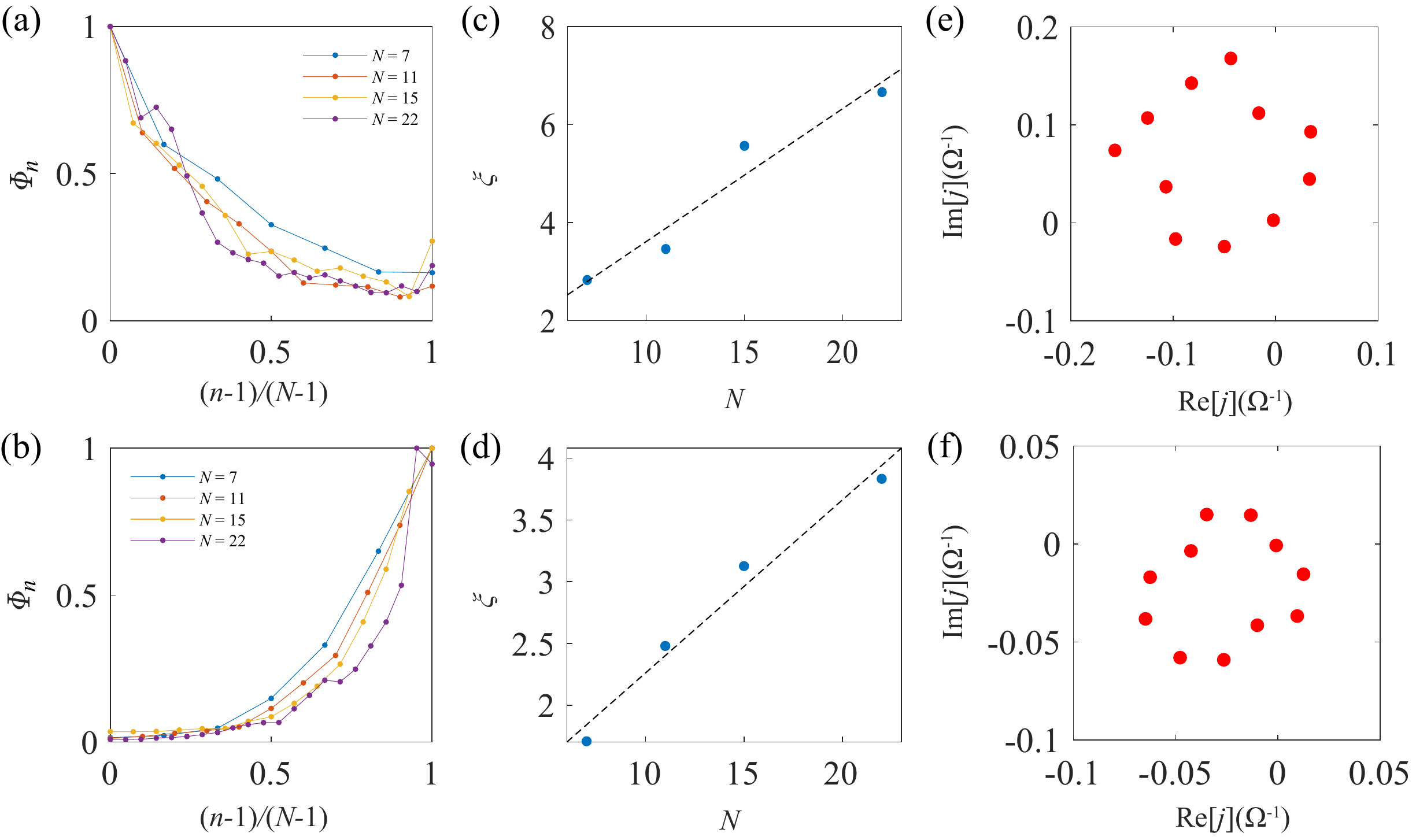}
	\caption{ Experimentally measured scale-free localization in electrical circuit. (a, b)   Normalized spatial distribution $\Phi_n$ of  the voltage at different normalized node index $(n-1)/(N-1)$ for the different lattice size $N$, where the node index  is mapped to the range $[0,~1]$. Here, (a) for  $C_1 = 9.4$nF and $C_v = 47$nF, and (b) for  $C_1 = 22$nF and $C_v = 2.2$nF.  (c, d) Localization length $\xi$ (blue dots) of bulk modes as a function of the lattice size $N$. The black dashed line denotes a linear fit to $\xi$.  The measured eigenvalues of the admittance (e) for $C_1 = 9.4$nF and $C_v = 47$nF, and (f) for  $C_1 = 22$nF and $C_v = 2.2$nF. }\label{fig4}
\end{figure*}

Figure  \ref{fig2}(c,d) shows the  normalized  spatial distribution $\Phi_n$ of  the peak voltage as a function of the normalized node index $(n-1)/(N-1)$ for the left- and right-localized skin modes at  the different lattice size $N$, where the node index  is mapped to the range $[0,~1]$. The state distributions at the different size are collapsed close to each other, indicating the size-dependent localization length. By exponentially fitting the state distribution, we extract the    localization length $\xi$ at different lattice size $N$ [see blue dots in Fig.~\ref{fig2}(e,f)]. After linearly fitting these dots, the localization length $\xi$ exhibits the linear dependence on the lattice size $N$. This indicates the existence of the scale-free localization for the skin modes controlled by the single non-Hermitian impurity.

\section{Experimental results of electrical circuits}
 
Our main results are the experimental verification of the scale-free localization induced by the single non-Hermitian impurity \cite{PhysRevResearch.5.033058} using the electrical circuit. The electrical-circuit network and fabricated experimental circuit board are shown in Fig.~\ref{fig1}(c,d). As shown in  Fig.~\ref{fig1}(c), two nodes within the circuit are interconnected via capacitors $C_1$ and  INICs, where the INICs have the equivalent capacitance of $\pm C_2$ in opposite directions. The first and last nodes are  connected through distinct capacitors and INICs, denoted as $C_v$ and $\pm C_\delta$, which serve as the single non-Hermitian impurity.  The parameters of the experimental electrical circuits are the same as ones used in the simulation. In addition to the random variations suffering from imperfect manufacturing processes, disorder is mainly introduced by the grounded capacitors  $C_{sn}$. The grounded capacitor $C_{sn}$ of each node is randomly chosen from a diverse set of  capacitors with capacitance ranging from $0$ to $C_S=10$ nF. The diagonal element of the circuit Laplacian  is given by   $ 1/(i\omega L_0)+i\omega(C_0+2C_1+ C_S/2)$,  and the circuit's reference frequency reads $f_0=(2\pi)^{-1} (C_0+2C_1+ C_S/2)^{-1/2}L_0^{-1/2}$, where the capacitance $C_1$ is  $C_1=9.4$ nF for the left-localized states, corresponding to $f_0 = 176 $ kHz, and it is $C_1=22$ nF for the right-localized states, corresponding to  $f_0 =164.5 $ kHz. A chirp signal spanning the frequency band from $0$ kHz to $250 $ kHz is used as the excitation. Details of the sample fabrication and experimental measurements are provided in the Appendix \ref{AppendixC}. 

The experimentally measured voltages of the admittance under the excitation of chirp signals are shown in Fig.~\ref{fig3}. We have experimentally designed two electrical circuits with different parameters of boundary capacitors acting as the single impurity, where the bulk parameters are fixed. To be specific, for the  $C_1 = 9.4$nF and $C_v = 47$nF, we plot the frequency-resolved voltage distribution [see Fig.~\ref{fig3}(a)]. The voltage is peaked around the frequency of $172.5 $ kHz, which matches well with the simulated result. By extracting the peak voltage at each node, its spatial distribution  is shown in Fig.~\ref{fig3}(b)], corresponding to the state distribution of non-Hermitian Hamiltonian $\mathcal{H}$ for the specific eigenvalue. This state is localized at the left side, indicating the occurrence of NHSE. While, for $C_1 = 22$nF and $C_v = 2.2$nF, the voltage is peaked around the frequency of $164 $ kHz [see Fig.~\ref{fig3}(b)], where we observe the right-side localized state [see Fig.~\ref{fig3}(d)]. The experimental results indicate the existence of the anomalous skin-mode localization   controlled by the single impurity in spite of the bulk hopping direction.

To verify the scale-free localization property, we measured the site-resolved peak voltages for different sizes, as shown in Fig.~\ref{fig4}(a,b). For the left-side skin modes, the parameter of the impurity for all the samples is set as $C_1 = 9.4$nF and $C_v = 47$nF, and for  the right-side skin modes,  it is $C_1 = 22$nF and $C_v = 2.2$nF. Figure \ref{fig4}(a) plots the peak voltages of different samples as a function of normalized node index $(n-1)/(N-1)$, where the node index is normalized to the range $[0,~1]$.  These left-side skin modes are not collapsed, indicating the absence of scaled localization for $C_1 = 9.4$nF and $C_v = 47$nF.  There is also absence of scaled localization for right-side skin modes with the single-impurity parameters set as $C_1 = 22$nF and $C_v = 2.2$nF. After performing a linear fit of the localization length at different sizes, we observe scale-free localization behavior. This size-dependence of the localization behavior exhibits a significant deviation from the NHSE, where the localization length remains consistent for    different system size $N$,  as predicted in Refs.~\cite{PhysRevResearch.5.033058,Li2021}. This unique phenomena of scale-free eigenstates are usually accompanied by the emergence of complex eigenspectrum\cite{Li2021}, which has also been presented in Fig.~\ref{fig4}(e,f).

\section{Conclusion}

In summary, we have experimentally observed the anomalous non-Hermitian skin effects with skin-mode localization directions controlled by a single non-Hermitian impurity in  non-Hermitian disordered electrical circuits. Furthermore,  anomalous skin modes are verified to show the scale-free localization induced by the single non-Hermitian impurity by measuring the size-dependent localization length.  Our experimental results have proved the theoretical proposal on the scale-free localization induced by the single non-Hermitian impurity.   In the future, it would be interesting to investigate scale-free localization in higher dimensions.

\begin{acknowledgments}
T.L. acknowledges the support from the Fundamental Research Funds for the Central Universities (Grant No.~2023ZYGXZR020), Introduced Innovative Team Project of Guangdong Pearl River Talents Program (Grant No. 2021ZT09Z109),  and the Startup Grant of South China University of Technology (Grant No.~20210012). W.B.J is supported by the National Natural Science Foundation of China (NSFC) (Grant No.~U21A2093).

\end{acknowledgments}

\appendix
\section{Circuit Laplacian}\label{AppendixA}

Linear circuit networks, composed of linear components, can be characterized by a series of time-dependent differential equations. After applying the Fourier transformation with respect to the time, these equations can be simplified into a set of algebraic equations in the frequency domain \cite{Lee2018}. In the frequency domain, the relation of current and voltage between two nodes can be written as
\begin{align}\label{eq1A}
	I_{jk}(\omega) = \frac{V_j(\omega) - V_k(\omega)}{Z_{jk}(\omega)},
\end{align}
where $Z_{jk}(\omega)$ is the impedance between node $j$ and node $k$, and the impedances of capacitor, inductor and resistor are $Z_C(\omega) = 1/i\omega C, Z_L(\omega) = i\omega L$ and $Z_R(\omega)=R$. According to Kirchhoff's current law, the sum of all currents   entering and
leaving a node equals zero. This indicates that the input current $I_j$ at the node $j$ equals the sum of the currents leaving node $j$.
\begin{align}\label{eq2A}
	I_j = \sum_{k} I_{jk}.
\end{align}

According to Eq.~(\ref{eq1A}) and Eq.~(\ref{eq2A}), we can derive the circuit Laplacian of the electrical circuit  in Fig.~\ref{fig1}(c,d).  Two nearest-neighbor nodes are connected  through capacitor with the capacitance $C_1\pm C_2$, and grounded by a capacitor $C_0,~C_{sn}$,  an inductance $L_0$ and a resistance $R_0$. Two nearest-neighbor nodes are also connected  through an INIC in parallel.  The INIC acts as a capacitance of $\pm C_2$ in two opposite directions.  The circuit equation of the $n$th ($n \neq 1, N$) node is written as,
\begin{align}
	I_n = &\left(i\omega C_0+i\omega C_{sn}+\frac{1}{i\omega L_{0}}+R_0 \right)V_n \nonumber \\ 
	& +i\omega(C_1+C_2 )(V_n-V_{n-1}) \nonumber \\ 
	& +i\omega(C_1-C_2 )(V_n-V_{n+1} ).
\end{align}

The first and last nodes are connected via the capacitor with the capacitance $C_{v} \pm C_{\delta}$, grounded by the capacitor $C_L,~C_{s1}$, an inductance $L_0$ and a resistance $R_0$. The circuit equations for the first and last nodes can be written as
\begin{align}
	I_1=&\left(i\omega C_L+i\omega C_{s1}+\frac{1}{i\omega L_{0}}+R_0 \right) V_1\nonumber \\&+i\omega(C_v+C_{\delta} )(V_1-V_{N}) \nonumber \\&+i\omega(C_1-C_2 )(V_1-V_{2}),
\end{align}
and 
\begin{align}
	I_N = &\left(i\omega C_R+i\omega C_{sN}+\frac{1}{i\omega L_{0}}+R_0 \right) V_N\nonumber \\&+i\omega(C_1+C_{2} )(V_N-V_{N-1})\nonumber \\& +i\omega(C_v-C_{\delta} )(V_N-V_{1} ).
\end{align}
\normalsize

In order to have the same on-site potential for all the nodes, we set   $C_L = C_1+C_2-C_v-C_{\delta}+C_0$ and $ C_R = C_1-C_2-C_v+C_{\delta}+C_0$. Then, we achieve the circuit Laplacian as 
\begin{widetext}
\begin{equation}\label{eqHA} 
	\begin{split}
		\small
		J = &i\omega\begin{pmatrix}
			C_{s1}-\frac{C_S}{2} -\varepsilon(\omega) & -C_1+C_2& 0 & \dots & 0& -C_v-C_{\delta} \\
			-C_1-C_2 & C_{s2}-\frac{C_S}{2} -\varepsilon(\omega) & -C_1+C_2 & \dots& 0 & 0 \\
			-0 & -C_1-C_2 & C_{s3}-\frac{C_S}{2} -\varepsilon(\omega) & \dots& 0 & 0 \\
			\vdots & \vdots& \vdots & \ddots & \vdots& \vdots \\
			0 & 0 & 0 & \dots & C_{sN-1}-\frac{C_S}{2} -\varepsilon(\omega) & -C_1+C_2 \\
			-C_v+C_{\delta} & 0 & 0 & \dots& -C_1-C_2& C_{sN}-\frac{C_S}{2} -\varepsilon(\omega)
		\end{pmatrix},
	\end{split}
\end{equation}
\end{widetext}
where $\varepsilon(\omega)=1/( \omega^2 L_0)-2C_1-C_0- C_S/2+i /(\omega R_0)$, and $C_{sn}$ is the grounded capacitance ranging from 0 to $C_S$, serving as disorder. 
If we set $\pm C_2-C_1 = t \pm \gamma$, $\pm C_\delta-C_v = v \pm \delta$, and $C_{sn}-C_S/2  = V_n$, $J$ and $\mathcal{H}$ can be related by $J=i\omega[\mathcal{H}-\varepsilon(\omega)]$.

\section{Details of experimental implementation}\label{AppendixC}

\subsection{Experimental setup }

\begin{figure}[!tb]
	\centering
	\includegraphics[width=9cm]{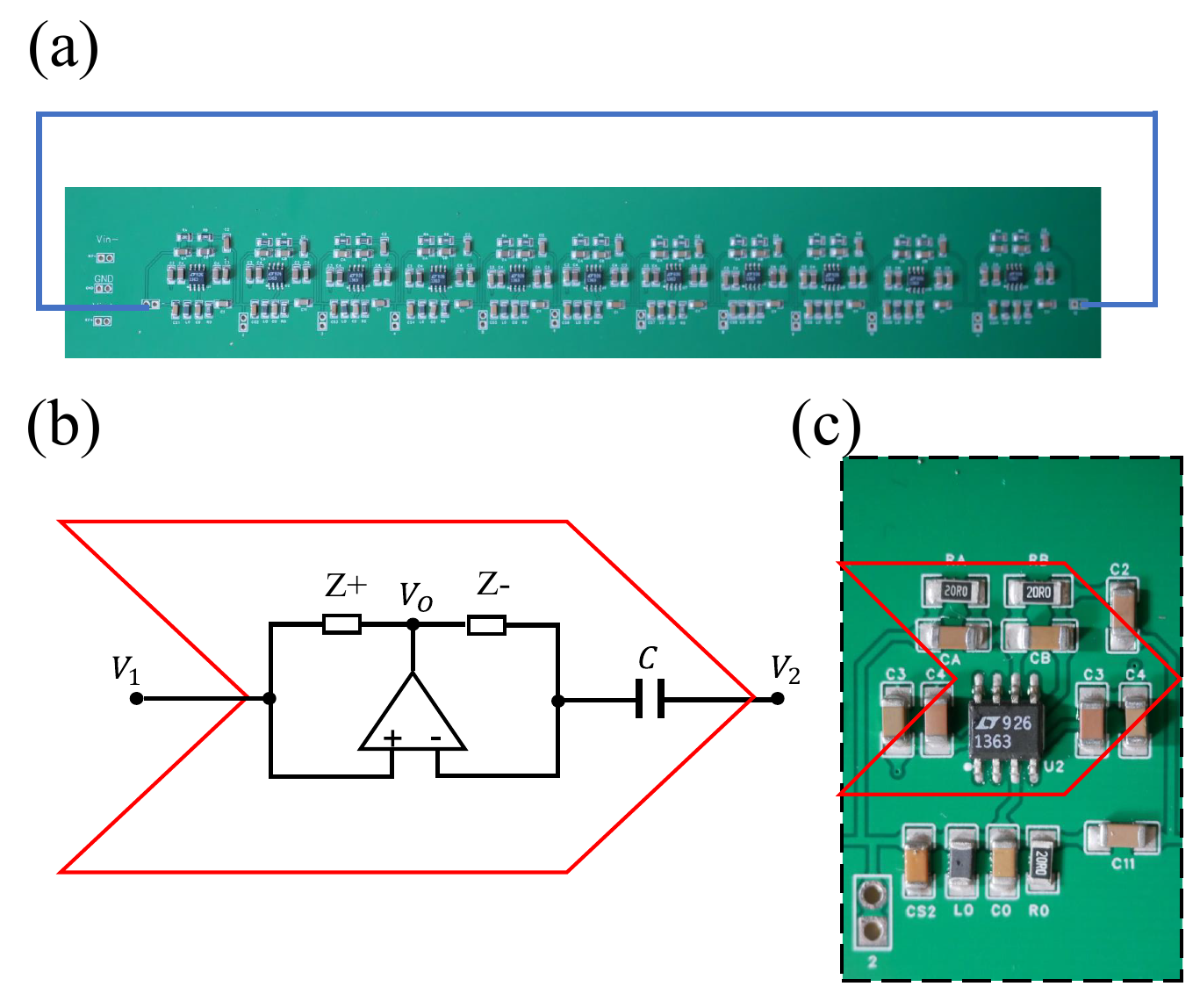}
	\caption{ (a) Experimental circuit board diagram containing eleven  unit cells for each board.  Multiple boards can be connected to create a longer chain. The first node and the last node are connected by the external wires acting as the single impurity. The nonreciprocal hopping between nodes $n$ and $n+1$ is realized by the negative impedance converters through current inversions (INICs) in  (b), where INIC  consists of capacitor, resistor and operational amplifier. (c) Circuit board  of each unit cell, where the red dashed curve indicates   the INIC.  }\label{fig3A}
\end{figure}

The disordered Hatano-Nelson model in the presence of a single non-Hermitian impurity is simulated by  the grounded circuit Laplacian $J$ of the electrical circuit in Fig.~\ref{fig1}(c). The fabricated circuit board is shown in Fig.~\ref{fig3A}(a), where the nonreciprocal hopping is realized via utilizing the impedance converters with current inversion (INIC) [see Fig.~\ref{fig3A}(b)]. The circuit board  of each unit cell is shown in Fig.~\ref{fig3A}(c), where the red dashed curve indicates the INIC. 

As shown in  Fig.~\ref{fig1}(c), two nodes within the circuit are interconnected via capacitors $C_1$ and  INICs, where the INICs exhibit equivalent capacitance  of $\pm C_2$ in opposite directions. The first and last nodes are  connected through distinct capacitors and INICs, denoted as $C_v$ and $\pm C_\delta$,which serve as the single impurity.  Each node is grounded by an inductor $L_0=4.7\mu $ H, a resistor $R_0$, and capacitors  $C_0=150$ nF and $C_{sn}$. In addition to the random variations suffering from imperfect manufacturing processes,  the random on-site potential is mainly realized by disordered grounded capacitors with the capacitance $C_{sn}$ of each node.   The value  of $C_{sn}$   is randomly selected  within  the range from $C_S=0$ nF to $C_S = 10$ nF.

The diagonal element of the circuit Laplacian  is given by   $ 1/(i\omega L_0)+i\omega(C_0+2C_1+ C_S/2)$, which vanishes at frequency  $f_0=(2\pi)^{-1} (C_0+2C_1+ C_S/2)^{-1/2}L_0^{-1/2}$.   In the experimental designs, the capacitance $C_1$ is chosen as $C_1=9.4$ nF for the left-localized states, corresponding to $f_0 = 176 $ kHz. It is $C_1=22$ nF for the right-localized states, corresponding to  $f_0 =164.5 $ kHz. They shows good consistency with the experimentally measured frequencies at the maximum voltage response, which were $172.5 $ kHz and $164 $ kHz, respectively.

For our specific experimental setup, a circuit board  with capacitors, inductors and resistors and operational amplifier was implemented on a Printed Circuit Board (PCB).  To ensure proper operation, the operational amplifier (LT1363) was powered by a DC current using the Keysight E3631A power supply. To reduce noise from the DC power supply, capacitors with capacitances of $2.2 $ $\mu$F and $1 $ $\mu$F were placed at the DC input of the operational amplifier. A chirp signal, covering a frequency range from $0$ kHz to 250 kHz, was generated using a Keysight 33500B waveform generator. The voltage source was interfaced with the Printed Circuit Board (PCB) through a shunt resistor $R=51$ $ \Omega$, functioning as the current input. A current source was input into the PCB to capture the voltage response using a Keysight DSOX4052A oscilloscope. A Fast Fourier Transform (FFT) was then applied to the measured time-domain voltage to obtain the voltage response in the frequency domain.

\subsection{Measurement}

In this section, we introduce the measurement of observables in  electrical circuits\cite{Helbig2020,PhysRevB.99.161114}.   In our experiment, we   measured the circuit's voltage response and complex admittance.  In following, we will discuss their correlation with the eigenvectors and eigenvalues of the Laplacian $J$.

\textit{Voltage response---} Based on $I=JV$, upon the input of the current, the voltage response at each node can be obtained by the inversion of Laplacian $J$.
\begin{align}\label{mastereqA}
	V = J^{-1} I = \sum_n \frac{1}{j_n} \ket{\psi_n^R} \bra{\psi_n^L} \ket{I},
\end{align}
where \(j_n\) is the $n$th eigenvalue of \(J\), and $\ket{\psi_n^R}$is the right eigenvector of \(J\) with eigenvalue \(j_n\) while $\bra{\psi_n^L}$ is the left eigenvector of \(J\). It indicates that all eigenstates contributes to voltage response, where  each eigenstate's contribution is weighted by its corresponding admittance \(j_n\) and   $\bra{\psi_n^L} \ket{I}$. The eigenvalue \(j_n(\omega)\) varies with frequency $\omega$. Assuming that, at the specific frequency, $J$ has the eigenvalue  very close to zero, the weight $ \bra{\psi_n^L} \ket{I} / j_n$ of the right eigenvector $\ket{\psi_n^R}$ can be considered significantly larger than contributions from the other right eigenvectors. In this case, the voltage response behaves as
	\begin{align}\label{mastereqA}
		V \sim\frac{\bra{\psi_n^L} \ket{I}}{j_n}\ket{\psi_n^R}.
	\end{align}
It indicates that, at this frequency, the voltage response $V$ of the circuit can be considered determined by the  eigenvector $\ket{\psi_n^R}$.

\textit{Complex admittance---} The voltage response depends on the inverse of the Laplacian $G=J^{-1}$  and input current. Their relationship can be expressed as
\begin{align}
	V = GI,
\end{align}
$G$ is the inverse of the Laplacian $J$. If we input a current at the single node $j$, the the voltage response:
\begin{align}
	V_i=G_{ij}I_j.
\end{align}
While we input a current at one node and measure the voltage response of all nodes, one can obtain a column of the matrix $G$. By repeating this process $N$ times, we obtain  $G$. Consequently, the Laplacian $J$ can be achieved by inverting $G$. The complex admittance is calculated using the reconstructed Laplacian $J$.

\end{document}